# Reliability of the *g* factor over time in Italian INVALSI data (2010-2022): What can achievement-*g* tell us about the Flynn effect?

Jakob Pietschnig[1], Sandra Oberleiter[1], Enrico Toffalini[2], & David Giofrè[3]

[1]Department of Developmental and Educational Psychology, Faculty of Psychology, University of Vienna, Austria
[2]Department of General Psychology, University of Padua, Italy
[3]Disfor, University of Genoa, Italy

**Abstract**

Generational intelligence test score gains over large parts of the 20th century have been observed to be negatively associated with psychometric g. Recent reports about changes in the cross-temporal IQ trajectory suggest that ability differentiation may be responsible for both changes in g as well as increasingly (sub-)domain-specific and inconsistent trajectories. Schooling is considered to be a main candidate cause for the Flynn effect, which suggests that school achievement might be expected to show similar cross-temporal developments. In the present study, we investigated evidence for cross-temporal changes in achievement-based g in a formal large-scale student assessment in Italy (i.e., the INVALSI assessment; N = 1,900,000+). Based on data of four school grades (i.e., grades 2, 5, 8, and 10) over 13 years (2010–2022), we observed little evidence for changes in achievement g in general. However, cross-temporal trajectories were differentiated according to school grade, indicating cross-temporal g decreases for lower grade students whilst changes for higher grade students were positive. These findings may be interpreted as tentative evidence for age-dependent achievement-g differentiation. The presently observed achievement g trajectory appears to be consistent with recently observed evidence for a potential stagnation or reversal of cognitive test score gains.

*Keywords*: Intelligence; Flynn effect; Achievement

**1. Introduction**

Generational IQ test score changes in the general population (i.e., the Flynn effect) were for the first time systematically documented in 1984 (Flynn, 1984). Despite a general positive trend on a global scale over most of the 20th century, these changes have been observed to be differentiated according to domain (i.e., fluid IQ gains are typically larger than crystallized gains) and yielded about 30, 35, and 25 IQ points increases in terms of full-scale, fluid, and crystallized intelligence globally from 1909 to 2013, respectively (Pietschnig & Voracek, 2015).

However, performance change trajectories were not linear over time and varied in strength across nations (Pietschnig & Voracek, 2015). Flynn effect patterns that have been observed in more recent years have become less consistent. Whilst in some countries Flynn effect decelerations were observed (e.g., USA; Rindermann & Thompson, 2013), others showed a stagnation, or a reversal (e.g., Denmark; Dutton et al., 2016). Notwithstanding, most of the available evidence on the Flynn effect so far had been framed on the taxonomy of Cattell's fluid and crystallized IQ distinction.

However, modern CHC model-based investigations (i.e., based on the Cattell-Horn-Carroll intelligence model; Schneider & McGrew, 2018) indicate that test score changes are differentiated according to more fine-grained stratum II CHC-based abilities (Lazaridis et al., 2022). Whilst some stratum II domains showed a positive Flynn effect (e.g., comprehension knowledge), others showed negative (e.g., working memory capacity) or ambiguous ones (e.g., fluid reasoning), whilst further domains were unaffected (e.g., processing speed).

However, one of the most striking and consistent results is the negative association of the Flynn effect with psychometric g (Pietschnig & Voracek, 2015). This seems paradoxical because predominantly cross-temporally increasing IQ (sub-)test scores appear to be incompatible with negative g associations. This can be thought of in terms of a decathlon. If performances on all ten events of the decathlon would be factor analyzed, a general factor (i.e., athletics-g) would emerge, because the performances on the subordinate factors (e.g., hurdling, shot putting) are intercorrelated.

If a decathlete were to train hurdling intensively, they would improve in this discipline, thus getting more points for their hurdling performance and their overall score. However, the performance in other subdisciplines would not increase, thus leading to a weakening of the positive athletics manifold. If we think about intelligence in an analogous manner, but with training taking place between generations instead of within individuals, increasing test scores that are negatively associated with psychometric g could be plausibly explained.

This has implications for population abilities, which may have undergone changes in terms of increasing specialization, thus conceivably manifesting itself in increasing ability differentiation. If this is the case, one should be able to observe a cross-temporally decreasing strength of the positive manifold. This idea is consistent with recent observations of increasingly differentiated Flynn effect patterns (Lazaridis et al., 2022). However, positive manifold strength changes have so far not been formally investigated.

Changes in the positive manifold are informative in regard to causes of the Flynn effect, regardless of the sign or strength of cross-temporal changes at a given time, because negative associations between g and the Flynn effect are unaffected by the direction and strength of changes. Understanding g-based changes is important, because cross-temporal decreases in g would suggest that (any) type of IQ changes are essentially shaped by increasing ability differentiation.

Here, we present the first formal assessment of cross-temporal changes in achievement g as a proxy for psychometric g. This is reasonable, because school achievement is a good proxy for intelligence (e.g., Pokropek et al., 2021). We assess cross-temporal g changes of g in population-representative samples of 2nd, 5th, 8th, and 10th graders in Italy (N = 1,900,000+) from 2010 to 2022 in large-scale formal educational assessment data from Italy (INVALSI; Istituto Nazionale per la VALutazione del Sistema di Istruzione).

## 2. Methods

### 2.1. Participants and materials

We analyzed data from the annual INVALSI assessments on population-representative Italian 2nd, 5th, 8th, and 10th grade student samples in mathematics and reading from 2010 to 2022. The INVALSI tasks are similar to PISA assessments and consist of 30–45 items, depending on survey year and grade. Test administration takes about two hours. Here, we analyzed data from 1,951,334 individual assessments (median n = 27,153 per year and grade) for math and reading (math subscales: "data and predictions", "numbers", "relations and functions", "space and figures", reading subscales: "grammar", "reading comprehension"; raw data is available upon request from the INVALSI institute).

### 2.2. Data analysis

To assess the meaningfulness of our data analyses, we first calculated the mean accuracies divided by sub-areas/types of items. Then, we performed confirmatory factor analyses to establish that scores were clustered according to the conceptually assumed reading and math areas. Consequently, we fitted a bifactor model to account for the expected data structure and facilitate g-factor-related examinations. Two (orthogonal) factors emerged, representing the g factor and a specific reading factor, modeling the two reading sub-areas (Fig. 1).

**Figure 1.**

*Bifactor model fitted across years and grades.*

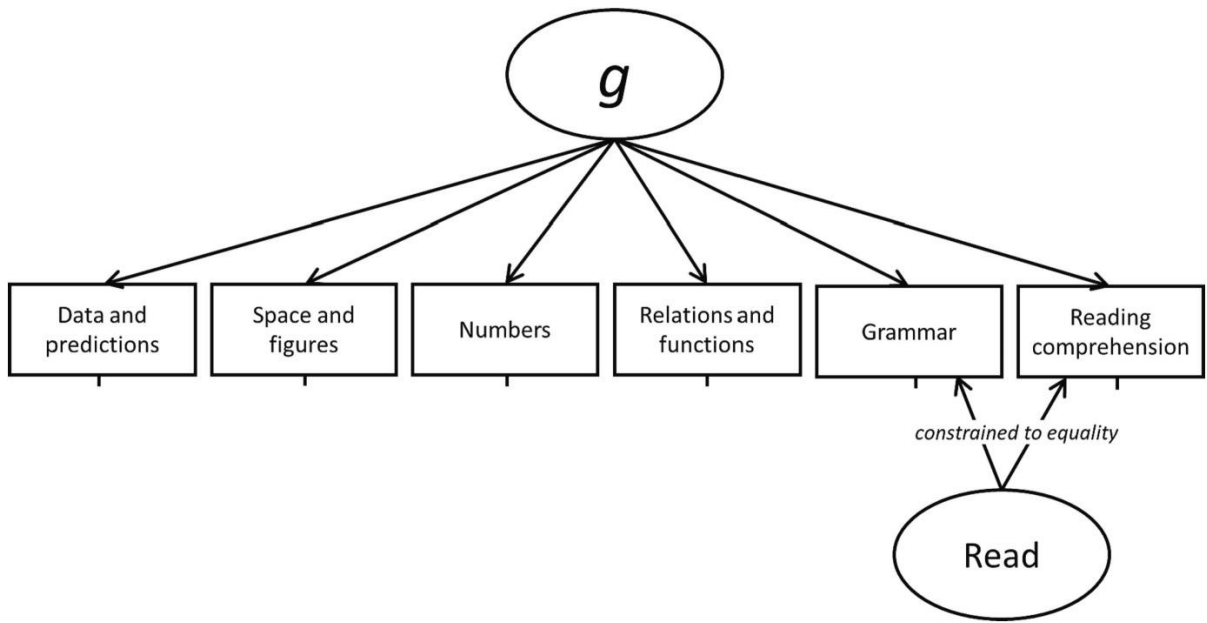

This approach allowed us to extract within-year i) model fit indices (RMSEA, SRMR, CFI, NNFI), ii) McDonald's omega ($\omega$) reliability indices for both latent factors, and iii) average explained variances (R2). Subsequently, we used generalized linear models to assess the effects of year, grade, and their interaction, on both $\omega$ and R2. We fitted a series of alternative models with all combinations of predictors on both indices of interest to assess model relevance using the Widely Applicable Information Criterion (WAIC; see Supplement S1 for a detailed description of our data analysis approach as well as specifics about our used R-packages). Secular test score changes are unavailable (such changes cannot be meaningfully reported for our data, because INVALSI data are restandardized within each assessment, thus rendering between-cohort comparisons uninformative for standard scores; due to potential changes of between-cohort item difficulties, raw score-based changes are uninformative).

## 3. Results

A total of 39 bifactor models were fitted, including 12 for grade 2, 12 for grade 5, 8 for grade 8, and 7 for grade 10 (data for year 2010 and since 2018 were unavailable). Fit indices were satisfactory: median *RMSEA* = 0.041 (range [<0.001, 0.092]); median *SRMR* = 0.010 (range [0.001, 0.030]); median *CFI* = 0.996 (range [0.978, 1.000]); median *NNFI* = 0.991 (range [0.958, 1.000]; see Supplementary Table S2 for details).

The best fitting model according to WAIC included both main effects and their interaction (Table 1; Fig. 2). R2 values were largely consistent, ranging between 0.40 and 0.60 across years, but with a slight tendency to increase at least in the latter grades (8 and 10). Interestingly, the sign of the main effect for year reversed when grade by year interactions were included, thus indicating decreases in achievement g over time, although the main effect did not reach significance. Results based on ω were virtually identical to $R^2$-based analyses, indicating the best fit for the model including the year by grade interaction (Table 1; Fig. 3).

**Figure 2**

*Model variance by achievement* g *as a function of grade and year. Shaded areas represent 95% Bayesian Credible bands.*

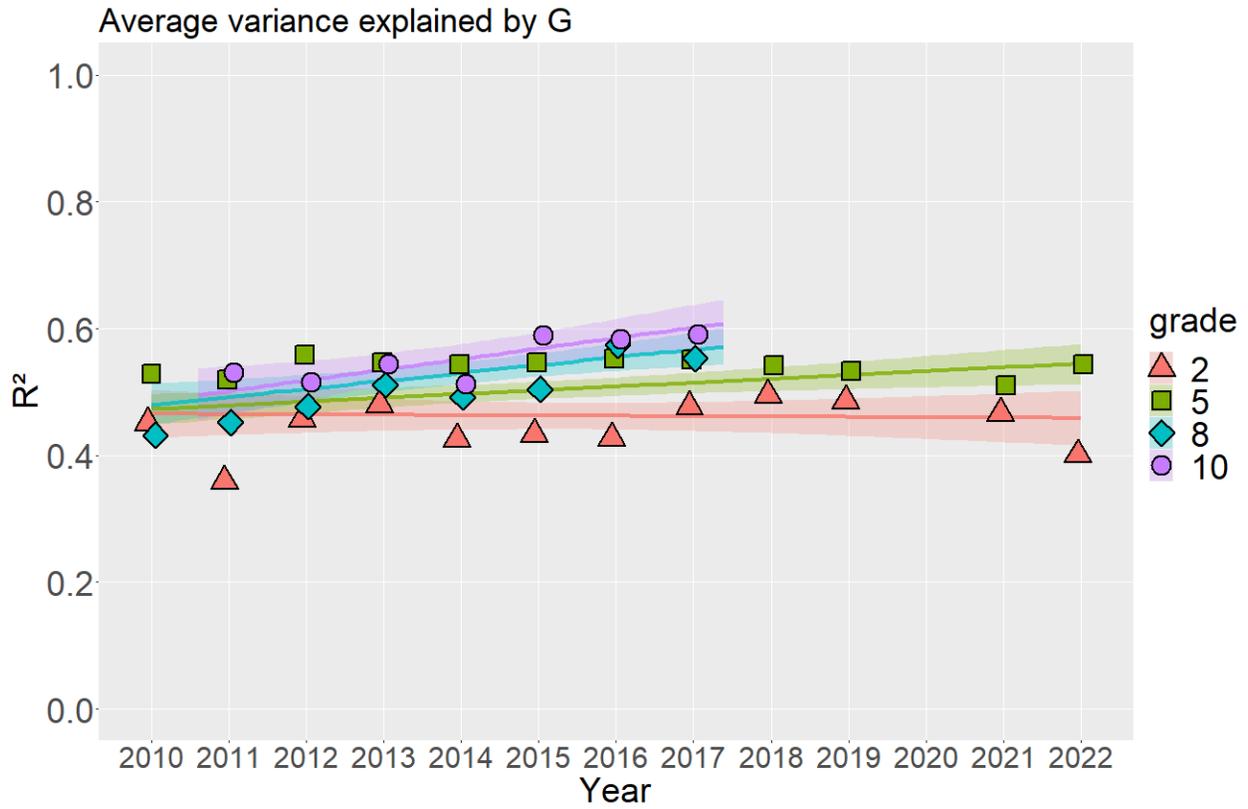

**Figure 3**

*Reliability (ω) of achievement g as a function of grade and year. Shaded areas represent 95% Bayesian Credible bands.*

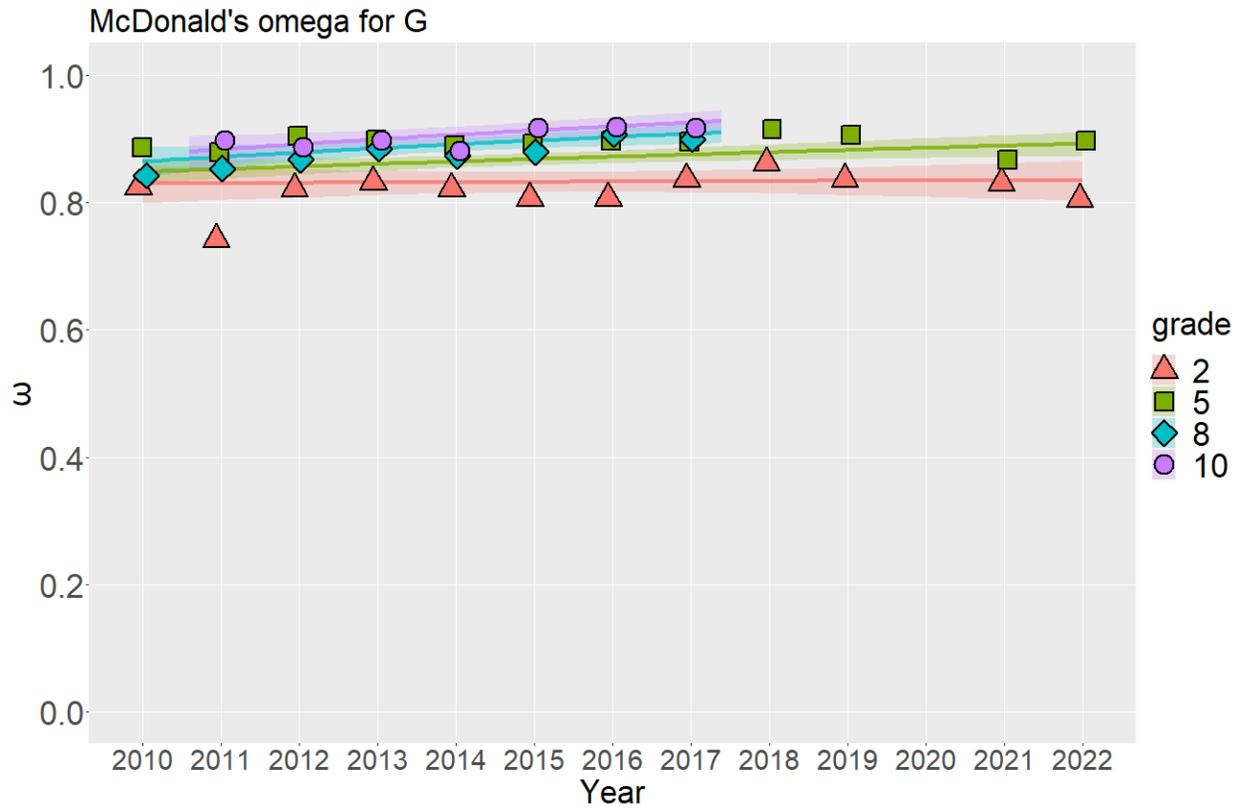

**Table 1**

Parameter estimates (95%CI) for $R^2$ and on $\omega$.

| Dependent variable | Model | Parameter | | | | WAIC |
|---|---|---|---|---|---|---|
| | | Intercept | Year | Grade | Year * Grade | |
| $R^2$ for g | | | | | | |
| | Intercept only | 0.020 (-0.051, 0.092) | - | - | - | -113.052 |
| | Year | -0.025 (-0.154, 0.107) | 0.009 (-0.013, 0.031) | - | - | -111.289 |
| | Grade | -0.217 (-0.349, -0.089) | - | 0.042 (0.022, 0.063) | - | -127.095 |
| | Year + Grade | -0.343 (-0.515, -0.173) | 0.019 (0.002, 0.036) | 0.048 (0.027, 0.068) | - | -129.884 |
| | Year * Grade | -0.158 (-0.366, 0.051) | -0.020 (-0.054, 0.015) | 0.010 (-0.025, 0.044) | 0.009 (0.002, 0.015) | -135.432 |
| $\omega$ for g | | | | | | |
| | Intercept only | 1.882 (1.771, 1.989) | - | - | - | -142.452 |
| | Year | 1.835 (1.645, 2.031) | 0.009 (-0.024, 0.044) | - | - | -140.581 |
| | Grade | 1.453 (1.287, 1.625) | - | 0.081 (0.052, 0.111) | - | -165.848 |
| | Year + Grade | 1.297 (1.089, 1.518) | 0.025 (0.002, 0.050) | 0.087 (0.059, 0.114) | - | -168.644 |
| | Year * Grade | 1.496 (1.207, 1.782) | -0.016 (-0.062, 0.032) | 0.045 (-0.003, 0.095) | 0.010 (0.000, 0.019) | -170.744 |

## 4. Discussion

We observed no substantial evidence for a weakening of g-related INVALSI-based achievement over time. Both explained variances as well as omega-based reliabilities showed non-significant decreases over time in our best-fitting models. Significant interactions between school grades and years indicated differentiated effects of elapsing time on the positive manifold strength.

Although the main effects of year on g were (non-significantly) negative for both $R^2$- and ω-based analyses, the significant interaction with grade suggested cross-temporal trajectory differences. In this vein, results for younger participants (i.e., those in lower grades) showed evidence for a decreasing strength of the positive manifold. Older students (i.e., those in higher grades) did not show decreases or even showed increases. Conceivably, these findings may be attributed to the lagged structure of the data.

Specifically, when shifting parameters of students in higher grades towards those years when they had been in second grade (i.e., thus artificially assuming identical ages of participants when they had been the same age as the youngest cohorts), a curvilinear pattern of initially increasing, then stagnating, and subsequently decreasing values would emerge. Such a pattern resembles the inverse u-shape of the Flynn effect in spatial task performance, as observed previously in Austria in a similar time frame (Pietschnig & Gittler, 2015). Similar curvilinear patterns have been found in Norway (Bratsberg & Rogeberg, 2018) and are consistent with findings from other countries (Dutton et al., 2016).

Curvilinear g-based trajectories may mean that changed Flynn effect patterns are rooted in increasing ability differentiation across cohorts. Conceivably, the presently observed interaction represents a manifestation of this very curvilinearity. Increases in ability differentiation are consistent with recent reports of distinct, non-monotonous Flynn effects in CHC model-based stratum II domains, which differed in terms of their signs (Lazaridis et al., 2022). Cross-temporal decreases in g are necessary consequences of diverging cross-temporal population IQ (sub-)domain trajectories.

However, it cannot be ruled out that the observed interaction of grade and year on g represents a consequence of a comparatively late onset of changes in achievement trajectories. Specifically, if g-related cohort-based changes were only to emerge in adolescence instead of childhood, then only higher-grade students would be expected to show cross-temporal change trajectories. This idea is consistent with evidence of no IQ test performance changes in preschoolers (Pietschnig et al., 2021). Conceivably, only comparatively old students (e.g.,10th graders) are subject to g-related changes which drives the presently observed interaction and indicates the cross-temporal stagnation or even increase of g. Notwithstanding, even if this were the case, the negative sign of the main effect for our best-fitting model contrasts this idea.

### 4.1. Limitations

First, it is possible that the cross-temporal g changes may be a consequence of periodical revisions of the INVALSI assessment (e.g., in terms of administration mode; Cornoldi et al., 2013) which may have led to unique variance component changes in different INVALSI achievement subtests. However, cross-temporal examination of the INVALSI bifactor structure did not indicate cross-temporal factor structure changes. Second, our observation period was comparatively short. However, our findings were virtually identical across two different analysis approaches, thus indicating stability of our findings. Third, it could be argued that the INVALSI subscales represent a rather narrow number of different abilities, thus possibly reducing the salience of the extracted g-factor. Whilst this idea cannot be outright dismissed, it should be noted that this makes g-related changes harder to detect, thus indicating that those changes that have been presently observed represent lower thresholds of the actual changes. Finally, we could not formally investigate test score changes due to within-region-, cohort-, and grade-based assessment revisions. However, although knowledge about potential test score changes in the INVALSI assessment would have been interesting, the interpretation of g-related changes does not depend on changes of scores.

## 5. Conclusions

In all, we present here evidence for cross-temporal stability of achievement g in a large-scale assessment of Italian students. Changes were differentiated according to school grade, indicating declines for younger students but potential increases of older ones. These school achievement-based findings appear to be consistent with the recently observed evidence for a potential stagnation or reversal of cognitive test score gains.

# References

Bundesärztekammer (2018). (*Muster-)Weiterbildungsordnung 2018* [*(Sample-)advanced trainings regulations*]. Bundesärztekammer.

Bratsberg, B., & Rogeberg, O. (2018). Flynn effect and its reversal are both environmentally caused. *PNAS, 115*, 6674-6678.

Cornoldi, C., Giofre, D., & Martini, A. (2013). Problems in deriving Italian regional differences in intelligence from 2009 PISA data. *Intelligence, 41*(1), 25-33. https://doi.org/10.1016/j.intell.2012.10.004

Diepgen, P., & Goerke, H. (1960). [*Brief summary table about the medical history*] [7th ed.]. Springer.

Dutton, E., van der Linden, D., & Lynn, R. (2016). The negative Flynn Effect: A systematic literature review. *Intelligence*, *59*, 163-169. https://doi.org/10.1016/j.intell.2016.10.002

Dworak, E. M., Revelle, W., & Condon, D. M. (2023). Looking for Flynn effects in a recent online U.S. adult sample: Examining shifts within the SAPA project. *Intelligence, 98*, 101734. https://doi.org/10.1016/j.intell.2023.101734

Flynn, J. R. (1984). The mean IQ of Americans: Massive gains 1932 to 1978. *Psychological Bulletin, 95*(1), 29-51. https://doi.org/10.1037/0033-2909.95.1.29

Merrifield, J. (2005). Specialization in a competitive education industry: Areas and impacts. *Cato Journal*, *25*, 317.

Lazaridis, A., Vetter, M., & Pietschnig, J. (2022). Domain-specificity of Flynn effects in the CHC-model: Stratum II test score changes in Germanophone samples (1996–2018). *Intelligence*, *95*, 101707. https://doi.org/10.1016/j.intell.2022.101707

Pietschnig, J. (2021). *Intelligenz: Wie klug sind wir wirklich?* [*Intelligence: How smart are we really?*]. Ecowing.

Pietschnig, J., & Gittler, G. (2015). A reversal of the Flynn effect for spatial perception in German-speaking countries: Evidence from a cross-temporal IRT-based meta-analysis (1977–2014). *Intelligence*, *53*, 145-153. https://doi.org/10.1016/j.intell.2015.10.004

Pietschnig, J., & Voracek, M. (2015). One Century of Global IQ Gains: A Formal Meta-Analysis of the Flynn Effect (1909–2013). *Perspectives on Psychological Science*, *10*(3), 282–306. https://doi.org/10.1177/1745691615577701

Pietschnig, J., Deimann, P., Hirschmann, N., & Kastner-Koller, U. (2021). The Flynn effect in Germanophone preschoolers (1996–2018): Small effects, erratic directions, and questionable interpretations. *Intelligence, 86*, 101544. https://doi.org/10.1016/j.intell.2021.101544Pokropek, A., Marks, G. N. & Borgonovi, F. (2021). How much do students' scores in PISA reflect general intelligence and how much do they reflect specific abilities? *Journal of Educational Psychology, 114*(5), 1121-1135. https://doi.org/10.1037/edu0000687

Rindermann, H., & Thompson, J. (2013). Ability rise in NAEP and narrowing ethnic gaps? *Intelligence*, *41*(6), 821-831. https://doi.org/10.1016/j.intell.2013.06.016

Schneider, W. J., & McGrew, K. S. (2018). The Cattell–Horn–Carroll theory of cognitive abilities. In D. P. Flanagan & E. M. McDonough (Eds.), *Contemporary intellectual assessment: Theories, tests, and issues* (pp. 73-163). The Guilford Press.

**Supplement S1**

**Data analysis**

For each child in each year and grade, we calculated the mean accuracies divided by sub-areas/types of items. Subsequently, confirmatory factor analyses were performed. Although we focused on the *g* factor in terms of our research question, we had to assess if item types were clustered according to the reading and math areas. Both to account for this data structure and to facilitate the calculation of the indices of interest (i.e., reliability of the *g* factor and average amount of observed performance variance explained by the *g* factor), we fitted a bifactor model. The model featured two (orthogonal) factors: the *g* factor and a specific reading factor that modelled the residual variance of the two reading sub-areas (because the sub-areas were only two for reading, their loadings were constrained to equality to facilitate convergence). Following the suggestions by Eid et al. (2017), we concluded that there was no need to add a specific "math" factor to model the residual variance for math sub-areas: this choice reflects the idea that math ability is a core aspect of general intelligence, and it facilitates model convergence without losing any fit with the data (see excellent fit indices in the results). Figure 1 illustrates the bifactor model fitted across years and grades.

*Figure 1*

Bifactor model fitted across years and grades

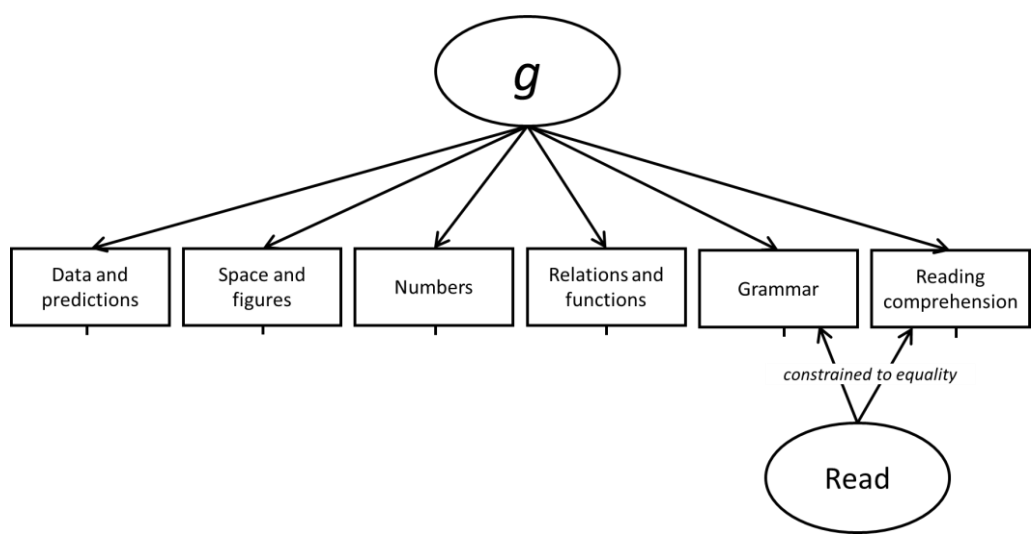

From the model in each year and grade, we extracted the following information: i) the fit indices (RMSEA, SRMR, CFI, NNFI); ii) the McDonald's omega ($\omega$) index of reliability for both latent factors; iii) the average variance explained ($R^2$) by each latent factor across the observed indicators (because the latent factors were constrained to being orthogonal by the bifactor model and thus shared no variance, the average variance explained could be calculated as the average squared standardized loadings of each factor).

Generalized linear models were used to assess the effects of year, grade, and their interaction, on both $\omega$ and $R^2$. Since both indices were continuous but constrained between 0 and 1 (and clearly presented no independence between their mean and variance), we modeled them using the beta distribution. Because this is not in the exponential family offered for the generalized linear models in R, we fitted the models using STAN via the interface implemented by the "brms" package of R. STAN is a probabilistic language written in C++ which allows the fitting of statistical models via MCMC algorithms for Bayesian inference, implementing a wide range of statistical distributions. Models are fitted with four chains each with 2,000 iterations (the first half are discarded as warmup, leaving with a total of 4,000 effective iterations per model). Uninformed default priors were used for all parameters. A Bayesian framework to model fitting was adopted only for ease of adopting the Beta distribution, which is not as easily implemented in other packages of R.

To assess the relevance of the effects of year, grade, and their interaction, we fitted a series of alternative models with all combinations of predictors on both indices of interest. Both predictors (year and grade were treated as continuous for parsimony and because the underlying dimensions, i.e., time and age, are naturally continuous). The best model was then selected using the Widely Applicable Information Criterion (WAIC) based on the posterior likelihood of the models. Better models have lower WAICs. In Bayesian inference, this is a widely used alternative to the AIC and BIC indices. Model point parameters were estimated as the median value of the posterior distribution, with 95% Bayesian Credible Intervals (BCI) calculated using the quantile method.

All analyses were conducted with the R free software (R Core Team, 2022), and the following packages: "lavaan" (Rosseel, 2012) for fitting CFA, "semTools" (Jorgensen et al., 2022) for extracting models reliabilities, "brms" (Burkner, 2017) for fitting generalized linear models via MCMC, "ggplot2" (Wickham, 2016) for data visualization.

**References**

Bürkner, P.-C. (2017). brms: An R Package for Bayesian Multilevel Models Using Stan. *Journal of Statistical Software, 80*(1), 1-28. doi:10.18637/jss.v080.i01

Eid, M., Geiser, C., Koch, T., & Heene, M. (2017). Anomalous results in *G* -factor models: Explanations and alternatives. *Psychological Methods, 22*(3), 541-562. https://doi:10.1037/met0000083

Jorgensen, T. D., Pornprasertmanit, S., Schoemann, A. M., & Rosseel, Y. (2022). semTools: Useful tools for structural equation modeling. R package version 0.5-6. Retrieved from https://CRAN.R-project.org/package=semTools

R Core Team (2022). R: A language and environment for statistical computing. R Foundation for Statistical Computing, Vienna, Austria. URL https://www.R-project.org/

Rosseel, Y. (2012). lavaan: An R Package for Structural Equation Modeling. *Journal of Statistical Software, 48*(2), 1-36. https://doi.org/10.18637/jss.v048.i02

Wickham, H. (2016). *ggplot2: Elegant Graphics for Data Analysis*. Springer-Verlag New York

**Supplement S2**

**Table S2**

Fit indices, $R^2$, and omega indices, for all available combinations of year and grade.

| Year | Grade | N | $\omega$G | $\omega$Read | $R^2$ G | $R^2$ Read | RMSEA | SRMR | CFI | NNFI |
|---|---|---|---|---|---|---|---|---|---|---|
| 2010 | 2 | 34,201 | 0.824 | 0.481 | 0.452 | 0.194 | 0.034 | 0.008 | 0.997 | 0.993 |
| 2011 | 2 | 30,928 | 0.742 | 0.453 | 0.359 | 0.203 | 0.000 | 0.001 | 1.000 | 1.000 |
| 2012 | 2 | 30,761 | 0.821 | 0.345 | 0.457 | 0.125 | 0.068 | 0.017 | 0.988 | 0.97 |
| 2013 | 2 | 24,194 | 0.832 | 0.302 | 0.480 | 0.114 | 0.015 | 0.004 | 0.999 | 0.999 |
| 2014 | 2 | 25,656 | 0.822 | 0.631 | 0.426 | 0.322 | 0.019 | 0.006 | 0.999 | 0.998 |
| 2015 | 2 | 20,408 | 0.807 | 0.432 | 0.433 | 0.192 | 0.044 | 0.012 | 0.995 | 0.987 |
| 2016 | 2 | 23,811 | 0.807 | 0.431 | 0.427 | 0.189 | 0.021 | 0.006 | 0.999 | 0.997 |
| 2017 | 2 | 24,067 | 0.836 | 0.409 | 0.477 | 0.183 | 0.039 | 0.009 | 0.996 | 0.991 |
| 2018 | 2 | 466,158 | 0.862 | 0.328 | 0.495 | 0.135 | 0.018 | 0.005 | 0.999 | 0.998 |
| 2019 | 2 | 22,711 | 0.836 | 0.316 | 0.486 | 0.126 | 0.038 | 0.009 | 0.997 | 0.992 |
| 2021 | 2 | 14,683 | 0.830 | 0.413 | 0.466 | 0.185 | 0.017 | 0.004 | 0.999 | 0.998 |
| 2022 | 2 | 14,902 | 0.805 | 0.357 | 0.401 | 0.162 | 0.026 | 0.010 | 0.997 | 0.994 |
| 2010 | 5 | 34,554 | 0.887 | 0.522 | 0.529 | 0.221 | 0.022 | 0.007 | 0.999 | 0.997 |
| 2011 | 5 | 30,832 | 0.879 | 0.437 | 0.520 | 0.166 | 0.019 | 0.005 | 0.999 | 0.998 |
| 2012 | 5 | 30,058 | 0.905 | 0.631 | 0.558 | 0.284 | 0.068 | 0.019 | 0.988 | 0.978 |
| 2013 | 5 | 23,962 | 0.898 | 0.597 | 0.546 | 0.260 | 0.064 | 0.018 | 0.989 | 0.980 |
| 2014 | 5 | 24,938 | 0.890 | 0.466 | 0.544 | 0.163 | 0.055 | 0.014 | 0.991 | 0.984 |
| 2015 | 5 | 20,542 | 0.892 | 0.471 | 0.546 | 0.184 | 0.019 | 0.005 | 0.999 | 0.998 |
| 2016 | 5 | 24,635 | 0.897 | 0.549 | 0.552 | 0.221 | 0.042 | 0.012 | 0.995 | 0.991 |
| 2017 | 5 | 24,523 | 0.895 | 0.527 | 0.551 | 0.197 | 0.055 | 0.014 | 0.992 | 0.985 |
| 2018 | 5 | 473,087 | 0.915 | 0.671 | 0.542 | 0.266 | 0.063 | 0.020 | 0.987 | 0.979 |
| 2019 | 5 | 23,862 | 0.906 | 0.627 | 0.533 | 0.202 | 0.054 | 0.018 | 0.990 | 0.984 |
| 2021 | 5 | 15,511 | 0.867 | 0.574 | 0.511 | 0.255 | 0.045 | 0.011 | 0.996 | 0.990 |
| 2022 | 5 | 15,511 | 0.897 | 0.579 | 0.543 | 0.285 | 0.061 | 0.019 | 0.990 | 0.982 |
| 2010 | 8 | 25,610 | 0.842 | 0.561 | 0.432 | 0.254 | 0.026 | 0.008 | 0.997 | 0.995 |
| 2011 | 8 | 25,892 | 0.853 | 0.614 | 0.452 | 0.247 | 0.041 | 0.014 | 0.994 | 0.989 |
| 2012 | 8 | 25,556 | 0.867 | 0.583 | 0.477 | 0.267 | 0.033 | 0.010 | 0.996 | 0.993 |
| 2013 | 8 | 28,153 | 0.884 | 0.617 | 0.511 | 0.267 | 0.045 | 0.012 | 0.994 | 0.988 |
| 2014 | 8 | 28,126 | 0.873 | 0.583 | 0.492 | 0.254 | 0.025 | 0.006 | 0.998 | 0.996 |
| 2015 | 8 | 28,523 | 0.879 | 0.606 | 0.504 | 0.257 | 0.030 | 0.007 | 0.997 | 0.995 |
| 2016 | 8 | 27,932 | 0.907 | 0.574 | 0.574 | 0.229 | 0.015 | 0.003 | 0.999 | 0.999 |
| 2017 | 8 | 28,003 | 0.899 | 0.558 | 0.553 | 0.239 | 0.043 | 0.010 | 0.995 | 0.991 |
| 2011 | 10 | 43,311 | 0.897 | 0.698 | 0.530 | 0.330 | 0.092 | 0.030 | 0.978 | 0.958 |
| 2012 | 10 | 41,654 | 0.887 | 0.619 | 0.516 | 0.290 | 0.071 | 0.018 | 0.985 | 0.972 |
| 2013 | 10 | 38,059 | 0.897 | 0.595 | 0.545 | 0.251 | 0.075 | 0.017 | 0.985 | 0.972 |
| 2014 | 10 | 36,895 | 0.881 | 0.559 | 0.513 | 0.242 | 0.057 | 0.015 | 0.990 | 0.982 |
| 2015 | 10 | 27,153 | 0.917 | 0.639 | 0.590 | 0.287 | 0.044 | 0.010 | 0.996 | 0.992 |
| 2016 | 10 | 33,857 | 0.919 | 0.712 | 0.583 | 0.336 | 0.064 | 0.016 | 0.991 | 0.983 |
| 2017 | 10 | 38,115 | 0.917 | 0.649 | 0.591 | 0.280 | 0.036 | 0.007 | 0.997 | 0.995 |